\documentclass[aps,physrev,twocolumn,superscriptaddress]{revtex4-2}

\usepackage{mhchem}
\usepackage{graphicx}
\usepackage{dcolumn}
\usepackage{bm}
\usepackage{hyperref}
\usepackage[mathlines]{lineno}
\usepackage{xcolor}
\usepackage{lipsum}
\usepackage{siunitx}
\usepackage{amsmath}
\usepackage[acronym]{glossaries}
\usepackage{amssymb}
\makeglossaries

\newcommand{\kBeq}{k_\text{B}}
\newcommand{\xpeq}{c_\text{p}}
\newcommand{\xneq}{c_\text{n}}
\newcommand{\T}{\vartheta}
\newcommand{\mupeq}{\mu_\text{p}}
\newcommand{\muneq}{\mu_\text{n}}

\newacronym{wo3}{WO$_3$}{tungsten oxide}
\newacronym{eios}{EIoS}{electrochemical ionic synapse}
\newacronym{icet}{ICET}{Ion-Coupled Electron Transfer}
\newacronym{mp}{MP}{Multi-Phase Polarization}
\newacronym{psg}{PSG}{phospho-silicate glass}
\newacronym{ecram}{ECRAM}{electrochemical RAM}

\begin{document}


\title{Theory of ultrafast conductance modulation in electrochemical protonic synapses by multiphase polarization}


\author{Michael L. Li}
\email[]{mli2360@mit.edu}
\affiliation{Department of Chemical Engineering, Massachusetts Institute of Technology, Cambridge, MA 02139, USA}

\author{Dingyu Shen}
\affiliation{Department of Electrical Engineering and Computer Science, Massachusetts Institute of Technology, Cambridge, MA 02139, USA}

\author{Jesus A. del Alamo}
\affiliation{Department of Electrical Engineering and Computer Science, Massachusetts Institute of Technology, Cambridge, MA 02139, USA}
\affiliation{Microsystems Technology Laboratories, Massachusetts Institute of Technology, Cambridge, MA 02139, USA}

\author{Martin Z. Bazant}
\email[]{bazant@mit.edu}
\affiliation{Department of Chemical Engineering, Massachusetts Institute of Technology, Cambridge, MA 02139, USA}
\affiliation{Department of Mathematics, Massachusetts Institute of Technology, Cambridge, MA 02139, USA}

\date{\today}

\begin{abstract}
Three-terminal electrochemical ionic synapses (EIoS) have recently attracted interest for in-memory computing applications. 
These devices utilize electrochemical ion intercalation to modulate the ion concentration in the channel material. The electrical conductance, which is concentration dependent, can be read separately and mapped to a memory state.
The state dependence on a non-volatile species leads to a desirable permanency in their memory.
To compete with alternative random access memory technologies, linear and symmetric conductance modulation is often sought after, properties typically thought to be limited by the slow ion diffusion timescale.
A recent study by Onen et al.\cite{onen_nanosecond_2022} examining protonic EIoS with a tungsten oxide (\ce{WO3}) channel revealed that this limiting timescale seemed irrelevant, and linear conductance modulation was achieved over nanosecond timescales, much faster than the \ce{WO3} bulk ion diffusion timescale.
This contrasts with previous studies that have showed similar conductance modulation with pulse timescales of milliseconds to seconds. 
Understanding the phenomena behind linear and symmetric conductance modulation in EIoS systems remains a crucial question gating technological improvements to these devices.
Here, we provide a theoretical explanation that demonstrates how linearity and symmetry arise from consistent control over the electrolyte-\ce{WO3} interface.
Between these studies, changes in the crystallinity of the \ce{WO3} channel were observed, affecting the material's thermodynamic properties and revealing that the device experiencing nanosecond pulse timescales undergoes phase separation.
We demonstrated that the electric field applied across the phase-separating polycrystalline \ce{WO3} channel polarizes the system and induces spontaneous phase separation, even when the bulk concentration remains in a stable single-phase region.
By coupling this with increased electron conductivity in the high-concentration filaments formed, the reaction environment at the gate electrode can be effectively controlled, resulting in ideal conductance modulation within the diffusion-limited regime.
Furthermore, to assist researchers in understanding how the thermodynamics of their synthesized materials influence EIoS device operation, we presented insights into how pulse timescales and relaxation timescales affect devices containing either solid-solution (amorphous) or phase-separating (polycrystalline) \ce{WO3}.
This work highlights the potential for phase-separating systems to overcome the traditional diffusion barriers that limit EIoS performance.

\end{abstract}

\maketitle

\newpage

\section{Introduction\label{sec:introduction}}
The scaling of the computation energy demands for AI and deep learning applications has garnered interest in in-memory computing.\cite{malik_governing_2013}
Ion-based electrochemical systems, specifically mixed ion-electron conductors, are popular material candidates for these devices because their electron conductivities can be tuned over multiple orders of magnitude through reversible ion intercalation.
Additionally, retention of the conductance state is also possible due to its dependence on a chemically inert species.
A promising architecture that leverages these materials is the \gls{eios}, inspired by biological neurons.\cite{kang_ion-driven_2022, talin_electrochemical_2025, huang_electrochemical_2023}
These devices, typically configured in a 3-terminal setup, utilize separate pathways to set and read the conductance state.
A key performance goal for successfully integrating these devices into novel in-memory computing systems is to achieve linear and symmetric conductance modulation with minimal switching times and energy consumption.\cite{yu_neuro-inspired_2018,jeong_memristor_2019,sebastian_memory_2020}
Traditionally, the switching time is considered to be confined to the ion diffusion timescale, $\tau_\text{diff} = L^2/D$.
Even if devices are reduced to nanometer thicknesses, the diffusion timescale for many ion-intercalation materials remains in the microsecond to millisecond range, far above the nanosecond switching times that would allow this technology to be most effective when combined with traditional computing architectures.\cite{tian_multiphase_2023}
\begin{figure*}
\includegraphics{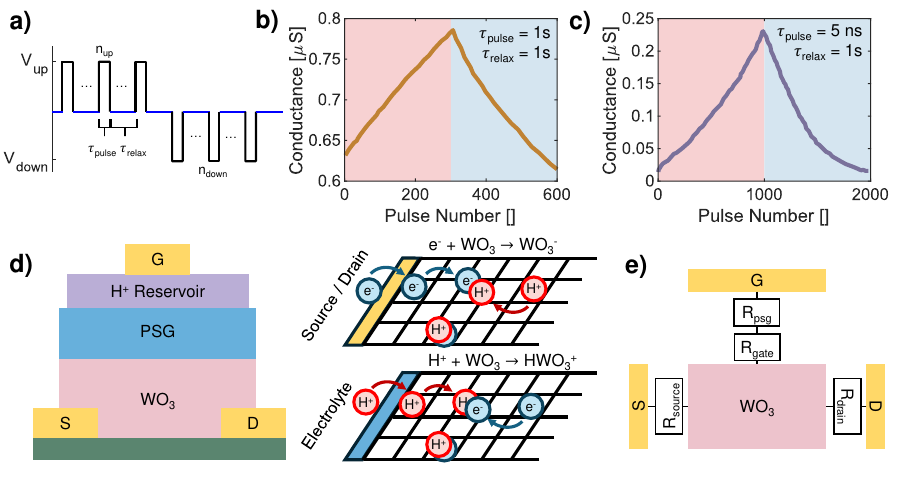}%
\caption{\label{fig:data}
Conductance modulation of protonic electrochemical ionic synapse with a \ce{WO3} channel. (a) Structure of pulse train experiments and results adapted from (b) study 1 \cite{onen_cmos-compatible_2021} and (c) study 2 \cite{onen_nanosecond_2022}.
Pulse train consists of consecutive trapezoidal pulses with a fixed applied potential and zero current relaxations.
(d) Schematic of three-terminal electrochemical ionic synapse and microscopic description of transfer reactions occurring at the gate, source, and drain electrodes.
(e) Simplifications of complex system geometry to rectangular \ce{WO3} bulk phase-field model domain and circuit model for electrode interfaces.
}
\end{figure*}
Recently, a few studies have achieved ultrafast switching speeds of 5 nanoseconds using an \gls{eios} with a tungsten oxide (\ce{WO3}) channel.\cite{onen_nanosecond_2022, tang_ecram_2018}
In these systems, ions originating from a reservoir gate material pass through a solid electrolyte and intercalate into the \ce{WO3} channel.
These devices also exhibited nearly linear and symmetric conductance modulation properties.
This combination of switching speeds and conductance modulation characteristics represents a significant advancement for \gls{eios} technologies toward meeting the performance benchmarks needed for analog non-volatile memories to be added to current computing architectures.\cite{xi2020memory}
Given the recency of these results, fundamental questions about the underlying phenomenon remain; answering these questions could significantly accelerate the future material design of \gls{eios} systems and, more broadly, ion-based electrochemical memory technologies.
Here, we focused on the protonic \gls{eios} system in Onen et al., as their study offered greater clarity on the material synthesis of \ce{WO3}.\cite{onen_nanosecond_2022}
Additionally, just one year earlier, Onen et al. demonstrated a protonic \gls{eios} device with a \ce{WO3} channel that could be switched much slower at 1 second speeds; correlating the switching speed and the underlying physical properties formed the basis of this work's results.\cite{onen_cmos-compatible_2021}
We developed a theoretical framework supported by phase-field modeling to reconcile the existence of linear and symmetric conductance modulation with switching speeds spanning eight orders of magnitude.
We showed that by designing \ce{WO3} to exhibit phase-separating thermodynamics, the local reaction environment for proton intercalation could be controlled.
This control forms the basis for achieving desirable conductance modulation even if the system is diffusion-limited.

\subsection{Experimental Studies and Results\label{sec:data}}
The data analyzed in this study are drawn from two experiments conducted by Onen et al. and published about a year apart. 
As shown in Fig.~\ref{fig:data}a-c, both studies employed trains of voltage pulses applied across the gate and source electrodes to drive the reversible intercalation of protons and electrons into \ce{WO3}. 
Between each pulse, zero-current relaxation holds were applied, and the conductance was measured independently through the source and drain electrodes. 
By altering the proton filling fraction in the \ce{WO3} lattice, previous studies have shown a 4-5 orders of magnitude change in electrical conductance.\cite{yao_protonic_2020, kamal_influence_2004, deb_optical_1973}
In practice, the studied devices operated within a low-filling regime, where a change of over an order of magnitude in \ce{WO3} conductance could be achieved.

The operating conditions of these experiments varied in pulse duration, which spanned eight orders of magnitude, from $\tau_{\text{pulse},1} = 1 \si{\second}$ to $\tau_{\text{pulse},2} = 5 \si{\nano\second}$. Throughout this paper, we refer to the system with slower switching speeds as study 1 \cite{onen_cmos-compatible_2021} and the system with ultrafast switching speeds as study 2 \cite{onen_nanosecond_2022}. Additionally, the applied voltages were significantly increased in magnitude from $V_\text{up,1} = 3$\si{\volt}, $V_\text{down,1} = -3$\si{\volt} in study 1 to $V_\text{up,2} = 11$\si{\volt}, $V_\text{down,2} = -8$\si{\volt} in study 2.
Because the magnitude and duration of the driving force were consistent across each pulse train segment, it is expected that the reaction environment's dynamics remained the same between pulses; otherwise, the linearity of the conductance modulation would not be anticipated.
Between these studies, the device size shrunk from a channel length of $L_1 = 25$ \si{\micro\metre} to $L_2 = 30$ \si{\nano\metre}.
Assuming that the limiting timescale for switching in these devices is ion diffusion, the timescale ratio between these studies was derived, $\tau_\text{diff,1}/\tau_\text{diff,2} = 6.94 \times 10^{5}$.
This ratio is three orders of magnitude lower than the experimentally observed ratio of pulse durations. This discrepancy strongly suggests that the limiting factor behind the switching speed of the conductance is not solely related to ion diffusion.

\section{Model\label{sec:model}}
To explain these experimental observations across different timescales, we developed a phase-field model that captures the electrochemical and transport phenomena in the \gls{eios}.
This model integrates thermodynamic and kinetic features of proton and electron intercalation under the influence of applied voltages and accounts for microstructural differences observed in amorphous and crystalline \ce{WO3} systems through their transport and thermodynamic parameters.
Unlike conventional battery and electrochromic systems, where electron and ion transfer typically occur at the same interface, the three-terminal architecture enabled a spatially decoupling of the intercalation of protons at the gate and electrons at the source and drain electrodes. \cite{deb_opportunities_2008, yu_electrochemical_1998}
Each intercalation reaction was described by a single-charge transfer reaction, driven by the externally applied voltage, as illustrated in Figure~\ref{fig:data}d.
\begin{subequations}
\label{eq:rxn_exp}
    \begin{gather}
        \mathrm{H}^+ \mathrm{(res)} + \mathrm{WO}_3 \rightarrow \mathrm{HWO}_3^+ \\
        \mathrm{e}^- \mathrm{(res)} + \mathrm{WO}_3 \rightarrow \mathrm{WO}_3^-
    \end{gather}
\end{subequations}
The resulting intercalated species, HWO$_3^+$ and WO$_3^-$, are assumed to occupy distinct lattices but are co-located due to the electrostatic interaction across lattices.
These species were treated as dopants in this model, where the proton or electron is allowed to transfer between adjacent \ce{WO3} sites through a hopping mechanism, $\mathrm{HWO}_3^+ + \ce{WO3} \rightarrow \ce{WO3} + \mathrm{HWO}_3^+$.
For the protons to enter the \ce{WO3}, they were transported through a phospho-silicate glass (PSG) solid electrolyte connecting the $\mathrm{PdH}_x$ reservoir and the \ce{WO3} channel.
This transport, as reported in previous works, was thought to be the dominant transport limitation.\cite{onen_cmos-compatible_2021}
In this work, we simplified the complex dynamics of the PSG electrolyte into a simple series resistance, $R_\text{psg}$.
The value of the resistance was found to be the largest in both the studied system and accounted for the majority of the applied potential drop between the gate and source electrodes.
Reducing this effective transport resistance would be critical in reducing the current energy costs to drive state-change in these devices.
Consequently, the local potential across the PSG-\ce{WO3} and \ce{WO3} bulk phase was approximately on the order of the thermal voltage scale, so the reaction current was modeled using the ion-coupled electron transfer theory.\cite{bazant_unified_2023}
\begin{subequations}
\label{eq:rxn_expr}
\begin{gather}
    j_\text{g} = j_\text{0,g} \xpeq^\alpha (c_\text{p,max}-\xpeq)^{1-\alpha} \left(e^{-\alpha \tilde{\eta}_\text{g}} - e^{(1-\alpha) \tilde{\eta}_\text{g}}\right) \\
    \kBeq \T \tilde{\eta}_\text{g} = \mu_\text{p,g} - \mu^\Theta_\text{\ce{WO3}} - e \Phi_\text{g} + J_\text{g} R_\text{l} \\
    j_\text{s/d} = j_\text{0,s/d} \left(e^{-\alpha \tilde{\eta}_\text{s/d}} - e^{(1-\alpha) \tilde{\eta}_\text{s/d}}\right) \\
    \kBeq \T \tilde{\eta}_\text{s,d} = \mu_\text{n,s/d} - \mu^\Theta_\text{\ce{WO3}} + e \Phi_\text{s/d}
\end{gather}
\end{subequations}
Here, $\tilde{\eta}$ denotes the dimensionless overpotential, $\Phi$ the macroscopic applied electrode potentials, and $j_0$ the exchange current density. 
In simplifying the dynamics of the transport through the PSG, all phenomena external to the \ce{WO3} bulk channel can be encapsulated within interfacial boundary conditions in Figure~\ref{fig:data}e.
\begin{figure}
\includegraphics{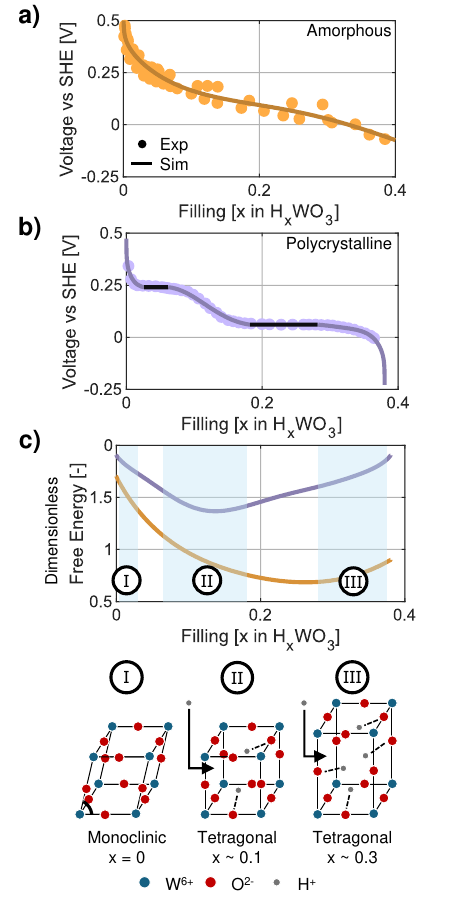}
\caption{\label{fig:mu}
Experimental and fitted open circuit voltage vs. SHE as a function of proton concentration for (a) amorphous \cite{crandall_theory_1976} and (b) polycrystalline \ce{WO3} \cite{jarman_electrochemical_1982}.
(c) Learned homogeneous free energy curve for the protons in amorphous and polycrystalline \ce{WO3}; the latter shows a multi-welled structure where comparisons have been made with literature-identified phases.\cite{henningsson_proton_2004}}
\end{figure}
Within the bulk \ce{WO3}, we adapted the multi-phase polarization theory developed by Tian and Bazant, which was used to investigate polarization phenomena in lithium titanate oxide undergoing resistive switching, to the system herein.\cite{tian_interfacial_2022, tian_multiphase_2023}
This theory extended the classical Nernst-Planck model for binary electrolyte to a phase-separating, lattice-constrained intercalation material.
Local electroneutrality was assumed and enforced via the relationship $\xpeq = \xneq = c$.
This assumption was supported by the small Debye length of approximately 0.4~\AA\cite{deb_optical_1973}, which was significantly smaller than the lattice constant of 5.2~\AA.\cite{jain_commentary_2013}
These species underwent transport through both electromotive and concentration gradient-driven fluxes, which are both rooted in the definition of the thermodynamic electrochemical potential for the two species.

\subsection{Thermodynamics}

From the details of device deposition from each study, the thermodynamics of each \ce{WO3} channel were determined to be distinct.
In study 2, an additional annealing step at 400\si{\celsius} was used to modify the \ce{WO3} structure.\cite{onen_nanosecond_2022}
Although the physical evidence of this change was not explicitly measured in that study, the annealing process has been extensively researched and is known to transform \ce{WO3} from an amorphous to a polycrystalline structure.\cite{zou_structural_2014, jafari_effect_2006}

The proton electrochemical potential utilized in this study was selected to address these differences, consisting of four components: (1) lattice entropy for a filling fraction (x in H$_\text{x}$WO$_3$, or equivalently $c/c_\text{n,max}$) between 0 and 0.38, (2) electrostatic interactions between protons and the local electric potential, (3) potential due to Cahn-Hilliard gradient penalty, and (4) concentration-dependent enthalpy of proton filling.
\begin{equation}
\label{eq:mup}
    \mupeq = \kBeq \T \ln{\left(\frac{\xpeq}{c_\text{p,max} - \xpeq}\right)} + \mu_\text{p,ex} - \kappa \nabla^2 \xpeq + e \phi
\end{equation}
The crystallinity-specific enthalpy was fitted to open-circuit voltage curves found in the literature, as displayed in Figure~\ref{fig:mu}a and b, using a Legendre polynomial series, $\mu_\text{p,ex} = \sum_n a_n P_n(c)$.
Doing so revealed thermodynamically unstable regions in the polycrystalline \ce{WO3} which led to voltage plateaus predicted at the binodal concentrations.
Fitting was accomplished using least squares regression method, where the simulated voltage contained the voltage plateaus derived from the underlying homogeneous potential.
In Figure~\ref{fig:mu}c, the derived homogeneous free energy for polycrystalline \ce{WO3} exhibited three characteristic local wells, corresponding to structural transitions from monoclinic to tetragonal crystal structures, as reported in the literature.\cite{henningsson_proton_2004}
The amorphous \ce{WO3}, which was found to exist in a single-phase regime across the entire concentration range, was modeled without the gradient penalty term.

Thermodynamics of the electrons were modeled using a conduction band filling model \cite{raistrick_thermodynamics_1981} parameterized by $\Delta \epsilon_\text{F}$ the change in Fermi energy level during conduction band filling, $d$ the dimensionality of the conduction band, and $\beta$ the fraction of inserted electrons that contribute to movement of the conduction band energy level.
\begin{equation}
    \muneq = \Delta \epsilon_\text{F} \left( \left(1 + \beta \frac{\xneq}{c_\text{n,max}}\right)^{2/d} - 1 \right) - e\phi
\end{equation}
It was assumed that these parameters were independent of the microstructure.

\subsection{Transport Dynamics}
The evolution of the concentration of protons and electrons was governed by their local conservation equations and the constraint of local electroneutrality.
\begin{subequations}
\label{eq:conservation}
    \begin{gather}
        \frac{\partial c}{\partial t} = \nabla \cdot \left( \frac{D_\text{p} c (c_\text{p,max}-c)}{c_\text{n,max} \kBeq \T} \nabla \mupeq \right) \\
        \frac{\partial c}{\partial t} = \nabla \cdot \left( \frac{D_\text{n} c_\text{n,max}}{\kBeq \T} \nabla \muneq \right)
    \end{gather}
\end{subequations}
Here, $D_\text{p}$ and $D_\text{n}$ are the diffusivities for protons and electrons, respectively.
Deviations from traditional flux expressions were derived from the definitions of species activities specific to this study.
To relate the parameters in this model with the known experimental electron conductance data \cite{yao_protonic_2020}, we established a constant proportionality relationship between these functions, $D_\text{n}(c)/\sigma(c) = \text{constant}$.
The boundary conditions for this system were determined by relating the species fluxes at the electrode-\ce{WO3} interfaces to the rates of intercalation reactions.
\begin{subequations}
    \label{eq:boundary-conditions}
    \begin{gather}
        \mathbf{n} \cdot \mathbf{F}_\text{p,g} = j_\text{g} / e \\
        \mathbf{n} \cdot \mathbf{F}_\text{n,g} = 0 \\
        \mathbf{n} \cdot \mathbf{F}_\text{p,s/d} = 0 \\
        \mathbf{n} \cdot \mathbf{F}_\text{n,s/d} = j_\text{s/d} / e
    \end{gather}
\end{subequations}
For polycrystalline \ce{WO3}, another set of boundary conditions were needed to capture the impact of the phase boundary on the surface free energy.
In this work, these considerations were neglected, leading to the natural boundary condition, $\mathbf{n} \cdot \nabla c_\text{p,g} = 0$.

\subsection{Implementation}
The model equations were implemented by scaling most parameters, such as the chemical and electrostatic potentials, with respect to their basic thermal scales.
The dynamical equations were transformed into a time-dimensionalized form in Equation~\ref{eq:dimensionless}.
Two critical process timescales arose from the rescaling of these governing equations.
\begin{subequations}
\label{eq:times}
    \begin{gather}
        \tau_\text{rxn} = \frac{j_0 L_\text{channel}}{e c_\text{p,max} D_\text{p}} \\
        \tau_\text{diff} = \frac{L_\text{channel}^2}{D_\text{p}}
    \end{gather}
\end{subequations}
The key parameters for the governing equations for amorphous and polycrystalline \ce{WO3} are shown in Table~\ref{tab:params}.
\begin{subequations}
    \label{eq:dimensionless}
    \begin{gather}
        \tilde{\mu}_\text{n} = \Delta \tilde{\epsilon}_\text{F} ( (1 + \beta \tilde{c})^{2/d} - 1) - \tilde{\phi} \\
        \tilde{\mu}_\text{p} = \ln{\left( \frac{\tilde{c}}{\gamma_\text{c,pn} - \tilde{c}} \right)} + \tilde{\mu}_\text{p,ex}(\tilde{c}) - \tilde{\kappa} \tilde{\nabla}^2 \tilde{c} + \tilde{\phi} \\
        \frac{\partial \tilde{c}}{\partial t} = \tilde{\nabla} \cdot \left(\frac{\tilde{c} (\gamma_\text{c,pn}-\tilde{c})}{\tau_\text{diff}} \tilde{\nabla} \tilde{\mu}_\text{p} \right)\\
        \frac{\partial \tilde{c}}{\partial t} = \tilde{\nabla} \cdot \left(\frac{\gamma_\text{d,np} \tilde{\sigma}(\tilde{c})}{\tau_\text{diff}} \tilde{\nabla} \tilde{\mu}_\text{n} \right)\\
        \left(\frac{\tau_\text{rxn,g}}{\tau_\text{diff}}\right) \mathbf{n} \cdot \tilde{\mathbf{F}}_\text{p,g} = \tilde{c}^\alpha (\gamma_\text{c,pn}-\tilde{c})^{1-\alpha} g(\tilde{\eta}_\text{g}) \\
        \tilde{\eta}_\text{g} = \tilde{\mu}_\text{p,g} - \tilde{\mu}^\Theta_\text{\ce{WO3}} - \tilde{\Phi}_\text{g} + \tilde{j}_\text{g} \tilde{R}_\text{l} \\
        \left(\frac{\tau_\text{rxn,s/d}}{\tau_\text{diff}}\right) \mathbf{n} \cdot \tilde{\mathbf{F}}_\text{n,s/d} = g(\tilde{\eta}_\text{s/d}) \\
        \tilde{\eta}_\text{s,d} = \tilde{\mu}_\text{n,s/d} - \tilde{\mu}^\Theta_\text{\ce{WO3}} + \tilde{\Phi}_\text{s/d} \\
        g(\tilde{\eta}) = e^{-\alpha \tilde{\eta}} - e^{(1-\alpha) \tilde{\eta}} 
    \end{gather}
\end{subequations}
Here, $\gamma_\text{c,pn} = c_\text{p,max} / c_\text{n,max}$ is the ratio of the maximum electron and proton concentrations, and $\gamma_\text{d,np} = D_\text{n,0} / D_\text{p,0}$ is the ratio of the electron and proton diffusivities. 
Finally, during the zero-current relaxation of these devices, the calculation of electron conductance was simplified by modeling electron transport through Ohm’s law, incorporating the time-dependent spatial concentration profile.
\begin{equation}
    \label{eq:ohm}
    \nabla \cdot (\sigma(c) \nabla \phi_s ) = 0
\end{equation}
The voltage difference across the source and drain for these calculations was 0.1 V.
The simulation is implemented using a finite volume method applied to a 2D rectangular mesh.
The sparse automatic differentiation package created by de la Gorce in MATLAB was used to speed up evaluation of the Jacobian for solving this model.\cite{delagorce2025sparse}
A semi-implicit time stepping where the convex splitting of the free energy is performed for numerical stability.
The results of the model are shown below.

\section{Results and Discussion}
\begin{figure*}
\includegraphics{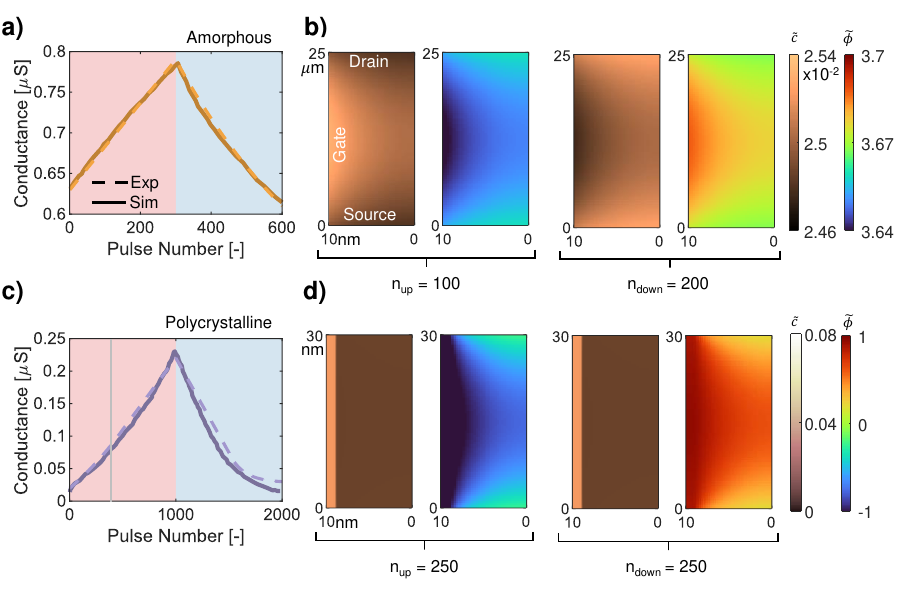}%
\caption{\label{fig:results} 
Simulated conductance modulation over pulse-train parameters used in (a) study 1 with amorphous \ce{WO3} and (c) study 2 with polycrystalline \ce{WO3}.\cite{onen_cmos-compatible_2021, onen_nanosecond_2022}
Gray line indicates when the system's average concentration reaches the material’s thermodynamic binodal concentration.
Key parameters used in the simulations are listed in Table~\ref{tab:params}. 
For (b) amorphous \ce{WO3} and (d) polycrystalline \ce{WO3} models, concentration and potential fields measured at the end of intermediate up and down pulses are displayed.
}
\end{figure*}
\setlength{\tabcolsep}{6pt} 
\begin{table}
\caption{\label{tab:params} Simulation parameters for amorphous and polycrystalline \ce{WO3}.}
\begin{tabular}{l@{}c@{}c}
Parameter & Value & Unit \\
\hline
$d$ & 3 & -- \\
$\Delta \epsilon_\text{F}$ & 3.45 & eV \\
$\beta$ & 1.0 & -- \\
$\gamma_\text{d,np}$ & 5 & -- \\
\hline
\multicolumn{3}{l}{Amorphous \ce{WO3}} \\
$\tau_\text{pulse}$ & $1.0$ & \si{\second} \\
$\tau_\text{diff}$ & $2.4 \times 10^{2}$ & \si{\second} \\
$\tau_\text{rxn}$ & $7.0 \times 10^{4}$ & \si{\second} \\
$\gamma_\text{c,pn}$ & 0.50 & -- \\
$\alpha$ & 0.5 & -- \\
$R_\text{psg}$ & 1.63 & \si{\ohm \metre \squared} \\
\hline
\multicolumn{3}{l}{Polycrystalline \ce{WO3}} \\
$\tau_\text{pulse}$ & $5.0 \times 10^{-9}$ & \si{\second} \\
$\tau_\text{diff}$ & $3.4 \times 10^{-4}$ & \si{\second} \\
$\tau_\text{rxn}$ & $3.2 \times 10^{-7}$ & \si{\second} \\
$\tilde{\kappa}$ & $1 \times 10^{-4}$ & -- \\
$\gamma_\text{c,pn}$ & 0.38 & -- \\
$\alpha$ & 0.4 & -- \\
$R_\text{psg}$ & 4.18 & \si{\ohm \metre \squared} \\
\hline
\end{tabular}
\end{table}
Our simulations successfully reproduced the conductance modulation in both amorphous (slow switching) and polycrystalline (fast switching) \ce{WO3} systems, as illustrated in Figure~\ref{fig:results}a and c. 
The relevant parameters used in each simulation are detailed in Table~\ref{tab:params}.
When observing the dynamics during the pulse shown in Figure~\ref{fig:results}b and d, both systems exhibited minimal variation in the concentration and potential at the electrolyte-\ce{WO3} interface.
Although there were significant differences in the values of local variables between the up and down pulse trains, only minor differences were observed between profiles selected during different pulses within a train.  
This highlights how the linearity of the conductance modulation depends on the consistency of the environment at the electrolyte-\ce{WO3} interface throughout consecutive pulses.
Upon examining the governing timescales, we found that these systems achieved this consistency under drastically different physical regimes.

In the case of amorphous \ce{WO3}, the dynamics were reaction-limited with the timescale relationship $\tau_\text{rxn} \gg \tau_\text{diff} \gg \tau_\text{pulse}$. 
As a result, most of the potential drop across the system occurred across the electrolyte-\ce{WO3} interface rather than the bulk \ce{WO3}.
It is important to note that this analysis neglects the potential drop across the phospho-silicate glass solid electrolyte, which accounts for nearly the entire macroscopic applied potential difference in both the amorphous and polycrystalline \ce{WO3} systems.
The small electric field across the bulk led to insignificant concentration polarization, as illustrated in Figure~\ref{fig:results}b.
In combination with the symmetry of the reaction model for the proton intercalation event, linear and symmetric conductance modulation was observed in the amorphous system.

In contrast, polycrystalline \ce{WO3} operated under diffusion limitations with the timescale relationship $\tau_\text{diff} \gg \tau_\text{rxn} \gg  \tau_\text{pulse}$. 
As a result, most of the potential drop across the system occurred over the bulk \ce{WO3} channel; the large electric field induced significant concentration polarization in the low-concentration bulk phase.
Locally, the proton filling fraction entered one of binodal gaps of polycrystalline \ce{WO3}, leading to formation of a high-concentration ($\tilde{c} \sim 0.1$) filamentary phase at the electrolyte-\ce{WO3} interface.
This region of thermodynamic instability is associated with a structural phase transition from monoclinic to tetragonal \ce{WO3}.
Due to the increased conductivity of this filament, which is an order of magnitude greater than that of the low-concentration phase, negligible electric fields are observed across the filament, as illustrated in Figure~\ref{fig:results}d.
\begin{widetext}
\begin{subequations}
    \label{eq:simplified}
    \begin{gather}
    \frac{\partial c}{\partial t} = \nabla \cdot\left(\frac{D_\text{p} c (c_\text{p,max} - c)}{c_\text{n,max} \kBeq \T} \left(\frac{\partial \mu_\text{total}}{\partial c}\right) \nabla c\right) \\
    \mu_\text{total} = \kBeq \T \ln{\left(\frac{c}{c_\text{p,max}-c}\right)} + \mu_\text{p,ex} + \Delta \epsilon_\text{F} \left( \left(1 + \beta \frac{\xneq}{c_\text{n,max}}\right)^{2/d} - 1 \right)
    \end{gather}
\end{subequations}    
\end{widetext}
This phenomenon can be understood mathematically, as the multiphase polarization theory in Equation \eqref{eq:conservation} simplifies to a single species Cahn-Hilliard model under the condition that $D_\text{n} \gg D_\text{p}$.
Consequently, the concentration and potential profiles along the gate were fixed throughout the pulse train.
As the pulses intercalated and de-intercalated protons into the bulk, the simulations showed that the filament size grew proportionally to the proton capacity.
The linearity observed in the conductance modulation is a result of the total conductance of the channel being a weighted sum of conductance contributions from each phase.
\[
\sigma(t) = \sigma_\text{low} (L_{\text{total}} - L_{\text{filament}}(t)) + \sigma_\text{high} L_{\text{filament}(t)}
\]
Symmetry of the conductance modulation was achieved in these simulations by utilizing an asymmetric reaction charge-transfer coefficient, which accounted for the different magnitude of the up and down pulses used experimentally.

During the pulsing of the polycrystalline \ce{WO3} system, the bulk average proton filling fraction initially existed outside the thermodynamic two-phase coexistence region but eventually entered it.
When the average concentration was outside this region, the system relaxed into a single-phase state.
During the initial applied up pulses, electric-field induced concentration polarization in the \ce{WO3} bulk led to metastable phase separation, resulting in linear conductance modulation.
During the final applied down pulses, the electric field across the \ce{WO3} bulk suppressed the formation of the filament, and local proton concentration depletion was observed near the gate.
This local depletion reduced the rate of the electrolyte-\ce{WO3} interfacial reaction, leading to non-linear conductance modulation.
The difference in behavior between the up and down pulses at low average proton concentrations caused a breakdown in symmetry of the conductance modulation in the tails of Figure~\ref{fig:results}c.

A key challenge with the results presented is reconciling the proton diffusion coefficient used in this study with values reported in the literature.
The coefficient used in this work was $2.6 \times 10^{-8}$ cm$^2$s$^{-1}$, which is on the order of $\mathcal{O}(10^2-10^3)$ times larger than the bulk values typically found in previous studies.\cite{randin_proton_1982, dickens_hydrogen_1981, matsuo_high_2025}.
We hypothesize that this discrepancy in the diffusion coefficient may arise from a few factors.
A significant buildup of protons in the solid electrolyte could be occurring, functioning as a pseudo-capacitor with a timescale comparable to the pulse duration.
This effect may lead to a slow self-discharge of \ce{WO3} during the relaxation phase, and a notable increase in the measured conductance of \ce{WO3} during the initial part of the relaxation.
To the authors' knowledge, observation of this phenomenon has not been documented in the literature.
Additionally, this material property would need to be found for both phospho-silicate glass for protonic EIoS and LIPON for lithionic EIoS to explain nanosecond conductance modulation in both systems.
Another possible explanation relates to the differences in proton diffusivity between the nanometer-scale films and their traditional bulk counterparts.
Even the presence of a small quantity of hydrated \ce{WO3} could significantly enhance proton diffusivity, which has been shown to increase by four orders of magnitude, as demonstrated by Dickens and others.\cite{dickens_hydrogen_1981, mitchell2022critical}
Although the diffusivity used in this study shows a significant discrepancy compared to established literature, applying these parameters results in a diffusion timescale of approximately 0.1 seconds.
This duration should not exhibit any response to a 5 nanosecond pulse. 
Further experimental investigations into proton conductivity are necessary to identify the correct model parameters.

\subsection{Understanding Fundamentals of Phenomenon}
\begin{figure}
\includegraphics{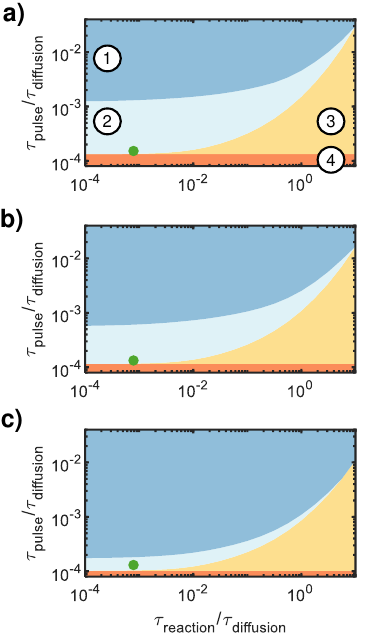}%
\caption{\label{fig:phase-diag} \textit{Kinetic} phase diagram resulting from variations of the ratio of pulse to diffusion timescales and the ratio of reaction to diffusion timescales for a phase-separating material.
Initial concentrations in simulations were set relative to the binodal concentration: (a) $c = 0.1 c_\text{bi}$, (b) $c = 0.25 c_\text{bi}$, and (c) $c = 0.50 c_\text{bi}$.
Four distinct phases were observed: (1) system reaches a stable two-phase regime; (2) system reaches a metastable two-phase regime that requires an electric field to maintain; (3) phase separation is not triggered by applying an electric field; (4) process timescale is shorter than the timescale for phase boundary formation. 
The location of the learned timescales parameterizing the experimental polycrystalline \ce{WO3} is shown in green, indicating metastable phase separation during pulsing.}
\end{figure}

To investigate this phenomenon further, we developed a simplified model in which the only variable dependent on the system is the thermodynamic model.
This allows for a direct comparison between amorphous, solid-solution \ce{WO3} and polycrystalline, phase-separating \ce{WO3}. 
In this model, system-dependent parameters from the experiments, such as the length scale, applied voltage, reaction kinetics, and the ratio of the electron to proton diffusivities, were fixed. 
A detailed overview of the parameters used in the model is provided in Table~\ref{tab:params2}.
\setlength{\tabcolsep}{6pt} 
\begin{table}
\caption{\label{tab:params2} Simplified model parameters for solid-solution and phase-separating materials.}
\begin{tabular}{l@{}c@{}c}
Parameter & Value & Unit \\
\hline
$V_\text{gate-source}$ & 0.128 & \si{\volt} \\
$\gamma_\text{d,np}$ & 10 & -- \\
$\alpha$ & 0.5 & -- \\
$R_\text{psg}$ & 0 & \si{\ohm \metre \squared} \\
\hline
\multicolumn{3}{l}{Solid-Solution} \\
$\Omega$ & 0 & \si{\joule} \\
$\tilde{\kappa}$ & 0 & -- \\
\hline
\multicolumn{3}{l}{Phase-Separating} \\
$\Omega$ & 4 & \si{\joule} \\
$\tilde{\kappa}$ & $1\times10^{-3}$ & -- \\
\hline
\end{tabular}
\end{table}
The thermodynamics of the system was simplified.
\begin{subequations}
\label{eq:new_thermo}
\begin{gather}
    \tilde{\mu}_\text{n} = -\tilde{\phi} \\
    \tilde{\mu}_\text{p} = \ln(\tilde{c}/(1-\tilde{c})) + \tilde{\Omega} \tilde{c} (1 - \tilde{c}) + \tilde{\phi} - \tilde{\kappa} \tilde{\nabla}^2 \tilde{c}
\end{gather}
\end{subequations}
The solid-solution system was parameterized with $\Omega = 0 \kBeq \T$ and $\tilde{\kappa} = 0$, whereas the phase-separating system was parameterized with $\Omega = 4 \kBeq \T$ and $\tilde{\kappa} = 1\times 10^{-3}$.
Additionally, the linear relationship between electron conductivity, from experiments \cite{yao_protonic_2020}, and electron diffusivity was utilized. 

Two key parameters govern the model: (1) the ratio of the reaction to diffusion timescales, denoted as $\tau_\text{rxn}/\tau_\text{diff}$, and (2) the ratio of the pulse to diffusion timescales, denoted as $\tau_\text{pulse}/\tau_\text{diff}$.
For the phase-separating system, we constructed a \textit{kinetic} phase diagram, shown in Figure~\ref{fig:phase-diag}, by independently varying the two timescale ratios across different starting concentrations.
These starting concentrations were defined with respect to the binodal concentration.
The resulting phase diagram revealed four distinct operating regimes.
\begin{figure*}
\includegraphics{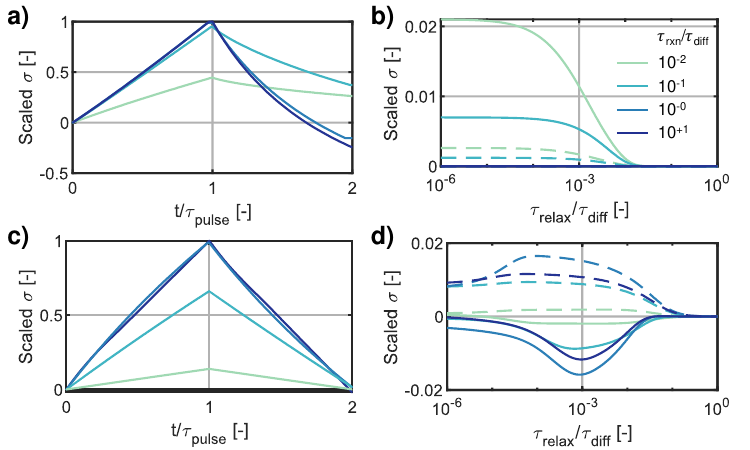}%
\caption{\label{fig:other}
Conductance modulation shown for varying $\tau_\text{rxn}/\tau_\text{diff}$ ratios, going from diffusion-limited to reaction-limited regimes, for models of (a) solid solution and (c) phase-separating materials.
Pulse trains with 100 intermediate relaxation steps and a total \textit{up} direction pulse time $\tau_\text{pulse}$ were used.
The ratio of the total pulse to the reaction timescale is fixed ($\tau_\text{pulse} / \tau_\text{rxn} = 0.1$).
Plotted time is scaled by the total pulse time, and conductance is scaled by the expression:  $(\sigma - \sigma_0) / (\max{\sigma}-\sigma_0)$.
Relaxation of material conductance during final up (solid) and down (dashed) pulse during pulse train for models of (b) solid solution and (d) phase-separating materials.
Plotted relaxation time is scaled by the diffusion timescale, and conductance is scaled by the following expression: $(\sigma - \sigma_\infty) / (\max{\sigma}-\sigma_0)$.
The starting concentration in both system models is $\tilde{c} = 0.05$, which lies in the two-phase coexistence region for the phase-separating system.}
\end{figure*}
Due to the experimental challenges in probing the microstructure of such small nanofilm devices, questions remain regarding how material design decisions could impact device performance.
\begin{enumerate}
    \item 
    Above a critical pulse timescale, the system intercalated enough protons during a single pulse so that the average concentration in the system lay within the two-phase coexistence region, resulting in phase separation of the bulk.
    \item
    System was diffusion limited $\tau_\text{diff} \gg \tau_\text{rxn}$, and pulses were short $\tau_\text{pulse} \ll 1$.
    The average proton concentration in \(\ce{WO3}\) stayed within the single-phase region.
    However, concentration polarization in the low concentration phase led to the formation of metastable filaments near the electrolyte-\ce{WO3} interface.
    \item 
    System was not severly diffusion limited $\tau_\text{diff} \ngtr \tau_\text{rxn}$, and pulses were short $\tau_\text{pulse} \ll 1$. 
    The average proton concentration in \ce{WO3} remained in the single-phase region; because the reaction timescale was large, the bulk \ce{WO3} does not exhibit large enough electric fields to induce local phase separation in the material.
    \item
    $\tau_\text{pulse}$ is much smaller than the $\tau_\text{diff}$ associated with a phase boundary. 
    The length of this phase boundary is given by $L_\text{boundary} = L_\text{channel} \sqrt{\tilde{\kappa}}$.
    The \ce{WO3} film does not have sufficient time to respond to the pulse.
\end{enumerate}

The parameters governing the polycrystalline \ce{WO3} system from Study 2 were compared against the \textit{kinetic} phase diagram, which indicated operation in Region (2) and was validated by the simulation results.

The model presented in this paper is utilized to examine how the \gls{eios} material thermodynamics affect conductance modulation properties when changes were made in the pulse and relaxation timescales.
In Figure~\ref{fig:other}a and c, conductance modulation across varying ratios of $\tau_\text{rxn} / \tau_\text{diff}$ were simulated.
For the solid solution model, decreasing this ratio, indicating increased diffusion limitation, resulted in increased concentration polarization, which ultimately limits the reaction as \(\tilde{c}\) approaches 1 or 0.
As the simulations started with a concentration closer to 0, the system experienced more limitations on the reaction kinetics during the down pulse train.
This led to a breakdown in the symmetry of conductance modulation, which became more pronounced as the timescale ratio decreased.
For the phase separating system model, the linearity and symmetry of the conductance modulation was found to be insensitive to the value of $\tau_\text{rxn} / \tau_\text{diff}$.

In Figure~\ref{fig:other}b and d, the conductance under various relaxation times was shown.
The solid solution system demonstrated a decrease in conductance as the system was allowed to relax, caused by the monotonic nonlinear concentration profile.
In contrast, the phase-separating system exhibited more intriguing relaxation behavior. 
This is due to the relaxation of the phase boundary occurring on a different timescale than the polarization phenomena, defined by the relationship $\tau_\text{boundary} / \tau_\text{diff} = \tilde{\kappa}$.
Notably, the change in conductance during the phase boundary relaxation displayed an opposite sign to the direction of the applied pulse.

Both trends highlight differences in the pulsing and relaxation dynamics that distinguish solid solution systems from those that undergo phase separation.
To the author's knowledge, these trends have not yet been confirmed by current literature. 
They represent identifiable features that demonstrate how material thermodynamics significantly affects the ability to achieve ideal symmetric linear conductance modulation in \gls{eios} technologies.


\section{Conclusions}

In this work, we investigated three-terminal EIoS systems with a \ce{WO3} channel showing linear and symmetric conductance modulation under pulse-train operation.
Despite an eight-order-of-magnitude difference in pulse timescales between two experimental studies, both systems displayed very similar conductance behaviors.
By focusing on the \ce{WO3} bulk response, we identified the material’s crystallinity as a key factor affecting its thermodynamic and dynamical responses across different proton filling fractions.

Using the multiphase polarization modeling framework, the symmetric modulation behavior in both systems was successfully reproduced.
However, further analysis of the simulation timescales revealed that the slow pulse time, amorphous \ce{WO3} system was reaction-limited, while the fast pulse time, polycrystalline \ce{WO3} was diffusion-limited.
The latter system seemed to contradict the traditional limitations placed on EIoS systems.
Simulations showed that the phase-separating thermodynamic nature of the polycrystalline \ce{WO3} system led to the formation of a secondary high-concentration filament at the gate during the application of an electric field.
Due to the increased conductivity of these films, further polarization was suppressed, resulting in a consistent electrolyte-\ce{WO3} interfacial environment over successive pulses.

To generalize these findings, we developed a simplified model to examine the fundamental aspects of the phenomenon in a thermodynamic phase-separating system.
It was shown that a certain range of process timescales (reaction, diffusion, and pulse) caused local filament formation at the gate, even when the bulk concentration was outside the two-phase coexistence region.
This model was extended to a simplified solid-solution material, allowing comparisons based on thermodynamic nature.
Given the limited experimental evidence for this mechanism in the literature, we examined how the shape of conductance modulation changed with variations in the ratio of reaction to diffusion timescale, $\tau_\text{rxn} / \tau_\text{diff}$.
Additionally, since phase boundary relaxation and concentration polarization relaxation occur on different timescales, significant differences in conductance relaxation profiles were observed. This is an important feature when considering these devices in computing systems, where conductance might be read before full relaxation.
These findings could assist researchers in determining the thermodynamic properties of their fabricated devices and fine-tuning them without depending on expensive characterization techniques.
Overall, our results highlight the potential of phase-separating materials to offer both precise control over interfacial reaction environments and faster response times in ion-based memory systems. 
Specifically, controlling the electrolyte-\ce{WO3} interfacial environment could be achieved when the system is diffusion-limited, leading to the experimental breakthroughs in device performance. 
These insights suggest that using phase-separating materials could be a promising approach to overcome traditional diffusion timescale limits in ion-based memories.

\section*{Data Availability}
The simulation results that support the findings of this article are openly available.\cite{li_2025_16637595}

\begin{acknowledgments}
    The authors thank Huanhuan Tian, Matthew Moy, and Mantao Huang for helpful discussions about the theory and the experimental setup. The authors thank Zhouhang (Amelia) Dai, Matthew Moy, and Debbie Zhuang for comments on the manuscript. This work was supported by Ericsson.
\end{acknowledgments}


%

\end{document}